\def\funp{{I\!\!P}}
\def\xp{x_{{I\!\!P}}}
\def\xpmin{x_{{I\!\!P}\mbox{\scriptsize{min}}}}
\def\xpmax{x_{{I\!\!P}\mbox{\scriptsize{max}}}}
\def\mrs{\mbox{\scriptsize{MRS}}}
\def\bl{\mbox{\scriptsize{BL}}}
\def\zmin{z_{\mbox{\scriptsize{min}}}}
\def\pt{p_{T}}
\def\ptt{p^{2}_{T}}
\def\kt{k_{T}}
\def\ktt{k_{T}^2}
\def\ktbar{k_{T}}

\def\kt{k_T}
\def\ktbar{\overline{k}_T}
\def\xbar{\overline{x}}
\def\xprime{x^{\prime}}
\def\xprimetwo{x^{\prime\prime}}
\def\zprime{z^{\prime}}
\def\xprimbar{\overline{x}^\prime}
\def\xprim2bar{\overline{x}^{\prime\prime}}
\def\ptbold{\mbox{\boldmath$p$}_T}
\def\ktbold{\mbox{\boldmath$k$}_T}
\def\ktboldbar{\mbox{\boldmath$\overline{k}$}_T}
\def\gapprox{\lower .7ex\hbox{$\;\stackrel{\textstyle >}{\sim}\;$}}
\def\lapprox{\lower .7ex\hbox{$\;\stackrel{\textstyle <}{\sim}\;$}}
\newcommand{\be}{\begin{equation}}
\newcommand{\ee}{\end{equation}}
\newcommand{\bea}{\begin{eqnarray}}
\newcommand{\eea}{\end{eqnarray}}
\def\gev{\mbox{\rm ~GeV}}
\documentstyle[11pt,epsfig,a4]{article}
\newlength{\dinwidth}
\newlength{\dinmargin}
\begin{document}
\titlepage
\begin{flushright}
DTP/98/12\\
March 1998
\end{flushright}

\vspace*{1in}
\begin{center}
{\Large \bf  Diffractive dijet photoproduction as \\
a probe of the off-diagonal gluon distribution}\\
\vspace*{0.5in}
K. \ Golec-Biernat\footnote{On leave from H.\ Niewodniczanski 
Institute of Nuclear Physics, 
Department of Theoretical Physics, ul. Radzikowskiego 152, Krakow, Poland.}, 
J.\ Kwiecinski$^1$, and A.D.\ Martin \\
Department of Physics, University of Durham, DH1 3LE, UK \\
\end{center}
\vspace*{2cm}
\vskip1cm 
\begin{abstract}
We propose exclusive diffractive dijet photoproduction as an ideal measure 
of the off-diagonal gluon distribution at high scales. 
We solve the off-diagonal evolution equations for the gluon and
quark singlet over the full kinematic domain. We discuss the nature 
of the solutions of these equations, which embody both DGLAP
and ERBL evolution.
We give predictions for the transverse momentum distribution of the jets.  
In particular we quantify the enhancement arising from the evolution 
of the off-diagonal parton distributions.
\end{abstract}

\newpage
\section{Introduction}
\label{sec:1}

\begin{figure}[tb]
   \vspace*{-1cm}
    \centerline{
     \epsfig{figure=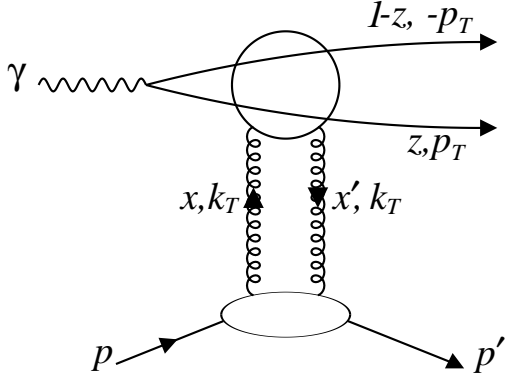,height=6cm}
               }
    \vspace*{-0.5cm}
     \caption{The QCD subprocess 
     $\gamma p \rightarrow (q\overline{q})\, p^{\prime}$, describing the exclusive
     diffractive photoproduction of dijets.}
\label{fig:0}
\end{figure}

Traditionally diffractive processes are described, within perturbative QCD, 
by two gluon exchange \cite{TRAD}.  An example is the $\gamma p \rightarrow q\bar{q}p$ 
process      
sketched in Fig.~\ref{fig:0}, where the outgoing $q\bar{q}$ system may emerge 
either as a      
vector meson or as two jets.  It has been argued that the amplitudes 
for these reactions      
are proportional to the conventional gluon 
distribution\footnote{The scale  at which the gluon 
distribution is   evaluated is $z(1-z)Q^2 + p_T^2 + m_q^2$, 
where $m_q$ and $\pm{\ptbold}$ 
are the mass and transverse momenta of the emitted    
quarks, and $z$ is the fraction of photon momentum carried by one of the 
quarks. So perturbative QCD is valid if
either $Q^2$ (with $z \sim 1/2$),  $\pt^2$ or $m_q^2$ is large.
} 
$G (\xp) \equiv \xp\, g (\xp)$, with $\xp$ given by     
\be
\label{eq:1}     
x_\funp \; = \; \frac{M^2 + Q^2}{W^2 + Q^2},     
\ee
where $M$ is the invariant mass of the diffractive $q\bar{q}$ system 
and $W$ is      
the $\gamma p$ centre-of-mass energy. 
The variable $\xp$ is the fraction of the proton
longitudinal momentum transferred to the diffractive
system by the exchange of the two gluons. 
A non-zero virtuality of the photon, $Q^2      
\neq 0$, allows electroproduction processes to be considered, 
in addition to photoproduction with $Q^2 = 0$.     
     
We notice that the two exchanged gluons in Fig.~\ref{fig:0} carry different momentum     
fractions $x$ and $x^\prime$, and so the process should actually 
be described by      
an off-diagonal\footnote{In the literature alternative nomenclatures are used
for ``off-diagonal'' distributions. In particular {\it non-forward, off-forward}
and {\it non-diagonal} are used in Refs.~\cite{RAD1}, \cite{JI1} and \cite{CFS}, 
respectively.
These works introduce different, but equivalent, definitions of the distributions.
Our work uses the formulation of \cite{RAD1}, but we prefer to retain the name 
``off-diagonal'' since we will be working in the forward direction.
} 
gluon distribution $G (x, x^\prime)$. For vector
 meson electro- or      
photo-production it turns out that it is a good approximation 
to use the off-diagonal      
distribution $G (x_\funp, 0)$ with $x = x_\funp$ and $x^\prime = 0$.  
The error made      in using the traditional diagonal distribution
$G(\xp, \xp) = \xp\, g(\xp)$ has been quantified
in Ref.~\cite{MR} (related work can be found in \cite{FS}).  
There the DGLAP-type evolution      
equation for the off-diagonal distribution
$G(x, x^\prime)$ was solved and compared      
with standard DGLAP evolution for $x\,g (x)$.  The ratio     
\be  
\label{eq:2}    
R (x, x^\prime) \; = \; \frac{G (x, x^\prime)}{x\,g(x)}     
\ee
was calculated as a function of $Q^2$, for different 
choices of (diagonal) starting      
distributions.  The ratio $R (x, x^\prime)$ was, 
as expected, found to be above unity     
and to increase with $Q^2$ and with $x_\funp \equiv x - x^\prime$.  
For example the     
off-diagonal effects for $J/\psi$ photoproduction can be estimated 
from the values    
obtained for $R$ at $x = x_\funp$ and $x^\prime = 0$.  
From Ref.~\cite{MR} we see    
that the $J/\psi$ amplitude is enhanced by
$R (x, 0) \simeq 1.1$ by off-diagonal    
evolution.  The prediction for the cross section is thus increased 
by a factor of about  $1.2$.     
     
Ref.~\cite{MR} considered only the off-diagonal effects arising 
from DGLAP-type evolution 
and so the analysis was restricted to the domain $x^\prime \geq      
0$.  The difference with the conventional diagonal distribution 
was found to be largest      
when $x^\prime = 0$ and to decrease rapidly with increasing $x^\prime$.  
That is $R      
(x, x^\prime)$ was largest when $x^\prime = 0$ and rapidly decreased 
to unity as      
$x^\prime \rightarrow x$.  However the DGLAP domain is only a part of a more      
general evolution of the off-diagonal distribution.  
In the region $x^\prime < 0$ the      
gluon $G (x, x^\prime)$ looks like the distribution {\it amplitude} 
for the proton to  
{\it emit} two gluons and consequently obeys an ERBL-type evolution equation  
\cite{ER,BL}.  Therefore processes which depend on a wider range of $x^\prime$,  
and in particular sample the $x^\prime < 0$ region, may be subject 
to much larger      
off-diagonal effects than the $J/\psi$ example that we mentioned above.  

One such  process is the exclusive diffractive production of a pair of (quark) 
jets with high values of
transverse momenta $\pm \mbox{\boldmath $p$}_T$.  For this process
it is necessary to work in terms of the off-diagonal gluon distribution $f$
unintegrated over the transverse momentum of the gluon,
\be
\label{eq:3}
f (x, \xprime, k_T^2) \; = \; 
\frac{\partial G(x,\xprime,k_T^2)}{\partial \ln k_T^2}\;.
\ee
The computation of the high $\pt$ dijet production cross section
requires integration over the entire region of $\kt$ (with an important
contribution coming from the region $\kt \sim \pt$). Hence this process
explores the detailed properties of the off-diagonal gluon distribution
(\ref{eq:3}) in a broad range of $\xprime$ and $\kt$.
In fact we will find that an important contribution comes 
from the region $|x^\prime| \sim \xp$ which should be
compared to  $\xprime = 0$ which is relevant for diffractive
$J/\psi$ production. As a result we are able to 
study the  much richer structure of 
the full evolution equations, which combine DGLAP-like  as well as ERBL-like 
features.

The content of the paper is as follows. In Section \ref{sec:2} we present
the formula for the diffractive dijet cross section driven by the off-diagonal
unintegrated gluon distribution. We carefully examine  the kinematic relations
relevant for the off-diagonal analysis. In Section \ref{sec:3} we present
a simplified analysis which provides valuable insight into the process, as well as
forming the basis for a comparison with our full off-diagonal treatment. The kinematic
relations relevant to the perturbative and nonperturbative regions are given is 
Section \ref{sec:4}. Then in Section \ref{sec:5} we study in detail the full
evolution equations of the off-diagonal distributions, which embody both DGLAP and ERBL
components. The detailed form of these equations is given in the Appendix. The effect
of using the off-diagonal distributions to predict the cross section 
for the exclusive diffractive dijet photoproduction is quantified in Section \ref{sec:6}. 
Finally Section \ref{sec:7} contains a brief summary of
our main results.
\section{General form of the diffractive dijet cross section}
\label{sec:2}

The differential cross sections for the exclusive diffractive production of dijets 
from transversely and longitudinally polarised photons,  
described by the QCD subprocess $\gamma^* p \rightarrow (q\overline{q})\,p^\prime$
shown in Fig.~\ref{fig:0}, 
have the following structure (see also \cite{NZ})
\be
\label{eq:4}
\left. \frac{d \sigma_{T,L}}{d^2 \ptbold \; dt}
\right|_{t = 0} 
= \frac{\alpha \alpha_S^2}{6 \pi}  \sum_q e_q^2
\int_0^1 dz 
\int \frac{d^2 \ktbold}{\kt^4}  \frac{d^2 \ktboldbar}{\ktbar^4} \;
\tilde{f}(x,\xprime,\kt^2) \;
\tilde{f}(\xbar,\xprimbar,\ktbar^2) \;
\Phi_{T,L}^q(z,\ktbold,\ktboldbar,\ptbold) \;, 
\ee
where $\pm \ptbold$ are the transverse momenta of the quark jets, $z$ and $(1-z)$
are their longitudinal momentum fractions, and $\ktbold$ is the transverse momentum
of the gluons.
The barred variables refer to the amplitude which is the complex conjugate  of 
the amplitude of the process shown in Fig.~\ref{fig:0}. Formula (\ref{eq:4}) was obtained
assuming that the imaginary part of the amplitude dominates,
with the  neglect of a  possible contribution from the real part. To compute
the imaginary part we have to take into account the four
possible ways that the two
exchanged gluons can couple to the two quarks forming the dijet system. This is reflected
both in the structure of the impact factors $\Phi^{q}_{T,L}$ and 
in the gluon distribution function $\tilde{f}$. Strictly speaking 
the notation for the arguments of $\tilde{f}$ is symbolic and
is only meant to indicate
that we are dealing with two gluons with different longitudinal momentum
fractions $x$ and $\xprime$. To be  precise $\tilde{f}$  is the following
linear combination of the off-diagonal distributions (\ref{eq:3}) 
\be
\label{eq:16a}
\tilde{f}(x, x^\prime, k_T^2) \;=\; \frac{1}{2}\;\biggl[
{f(x(\xprime), \xprime, k_T^2) 
+ f(x(\xprimetwo), \xprimetwo, k_T^2)}
\biggr]\;,
\ee
where the longitudinal momentum fractions  
($x(\xprime)~\mbox{\rm and}~x(\xprimetwo)$ 
of the first gluon, emitted from the proton in  Fig.~\ref{fig:0}, 
and $\xprime~\mbox{\rm and}~\xprimetwo$ of
the second gluon) depend on whether the second
gluon couples to the quark with momentum fraction $z$ or $(1-z)$. 
For these two configurations
we have respectively
\be
\label{eq:7}
x^{\prime} \; = \; 
\frac{k_T^2 + 2 \ptbold \cdot \ktbold}{z~(Q^2 + W^2)} 
~~~~~~\mbox{\rm and}~~~~~~
x^{\prime\prime} \; = \;
\frac{k_T^2 + 2 \ptbold \cdot \ktbold}{(1-z)~(Q^2 + W^2)}~, 
\ee
with
\be
\label{eq:5}
x(\xprime)\; = \; x_{\funp} + x^{\prime}~~~~~~~\mbox{\rm and}~~~~~~
x(\xprimetwo)\; = \; x_{\funp} + x^{\prime\prime}~,
\ee 
where $\xp$ is specified by (\ref{eq:1}) with a
diffractive  mass of the dijet system given by
\be
\label{eq:6}
M^2  = \frac{p_T^2 + m_q^2}{z\,(1 - z)}\;.
\ee
Notice that $\xp=\zeta\equiv x-\xprime$ (or $x-\xprimetwo$) plays the role
of the asymmetry variable for our process.  It is also important to note that
the momentum fractions of the
first gluon are always {\it positive}, whereas the fractions 
$\xprime$ and $\xprimetwo$ carried by the second gluon 
may be {\it negative} for
$\kt<2\pt$. In this case the second gluon is emitted, rather than
absorbed as  is shown in Fig.~\ref{fig:0}.
For the complex conjugate amplitudes we have analogous formulae 
for $\xprimbar$ and $\xbar^{\prime\prime}$ 
with $\ktbold$ replaced by $\ktboldbar$.

Finally, the impact factors $\Phi_T^q$ and $\Phi_L^q$ are \cite{NZ}
\bea
\label{eq:8}
\Phi_T^q (z,\mbox{\boldmath$k$}_T, \overline{\mbox{\boldmath$k$}}_T, 
\mbox{\boldmath$p$}_T) \; = \; [z^2 + (1 - z)^2 ]~ 
\Phi_1(z,\mbox{\boldmath$k$}_T, \overline{\mbox{\boldmath$k$}}_T, 
\mbox{\boldmath$p$}_T) + {m_q^2}~\Phi_2(z,\mbox{\boldmath$k$}_T, 
\overline{\mbox{\boldmath$k$}}_T, 
\mbox{\boldmath$p$}_T) 
\eea
and
\bea
\label{eq:9}
\Phi_L^q (z,\mbox{\boldmath$k$}_T, \overline{\mbox{\boldmath$k$}}_T, 
\mbox{\boldmath$p$}_T) \; = \; 
4\, Q^2 z^2\, (1 - z)^2~\Phi_2(z,\mbox{\boldmath$k$}_T, 
\overline{\mbox{\boldmath$k$}}_T, 
\mbox{\boldmath$p$}_T)~,
\eea
where
\bea
\label{eq:10}
\Phi_1 (z,\mbox{\boldmath$k$}_T, \overline{\mbox{\boldmath$k$}}_T, 
\mbox{\boldmath$p$}_T) \; = \; 
\left\{ \frac{ p_T^2 }
{[p_T^2 + \overline{Q}^2]^2} \: - \: \frac{
\mbox{\boldmath$p$}_T \cdot (\mbox{\boldmath$p$}_T + \mbox{\boldmath$k$}_T) 
}{[p_T^2 + \overline{Q}^2] [(\mbox{\boldmath$p$}_T + \mbox{\boldmath$k$}_T)^2
+ \overline{Q}^2]} \right. \nonumber \\
\\
\left. - \; \frac{
\mbox{\boldmath$p$}_T \cdot (\mbox{\boldmath$p$}_T + 
\overline{\mbox{\boldmath$k$}}_T)  
}{[p_T^2 + \overline{Q}^2] [(\mbox{\boldmath$p$}_T + 
\overline{\mbox{\boldmath$k$}}_T)^2
+ \overline{Q}^2]} 
+ \frac{(\mbox{\boldmath$p$}_T + 
\mbox{\boldmath$k$}_T) \cdot (\mbox{\boldmath$p$}_T + 
\overline{\mbox{\boldmath$k$}}_T)}{[(\mbox{\boldmath$p$}_T + 
\mbox{\boldmath$k$}_T)^2 + \overline{Q}^2] [(\mbox{\boldmath$p$}_T + \overline{
\mbox{\boldmath$k$}}_T)^2 + \overline{Q}^2]} 
\right\}, \nonumber
\\ \nonumber
\eea
and
\bea
\label{eq:11}
\Phi_2 (z,\mbox{\boldmath$k$}_T, \overline{\mbox{\boldmath$k$}}_T, 
\mbox{\boldmath$p$}_T) \; = \; 
\left\{ \frac{1}{[p_T^2 + \overline{Q}^2]^2} - \frac{1}{[p_T^2 + \overline{Q}^2] 
[(\mbox{\boldmath$p$}_T + \mbox{\boldmath$k$}_T)^2
+ \overline{Q}^2]} \right. \nonumber \\
\\
\left. - \; \frac{1}{[p_T^2 + \overline{Q}^2] 
[(\mbox{\boldmath$p$}_T + \overline{\mbox{\boldmath$k$}}_T)^2
+ \overline{Q}^2]} + \frac{1}{[(\mbox{\boldmath$p$}_T + 
\mbox{\boldmath$k$}_T)^2 + \overline{Q}^2] [(\mbox{\boldmath$p$}_T + \overline{
\mbox{\boldmath$k$}}_T)^2 + \overline{Q}^2]} \right\}, \nonumber
\\ \nonumber
\eea
with 
\be
\label{eq:12}
\overline{Q}^2 \equiv z\, (1 - z) \,Q^2 + m_q^2. 
\ee 
Cross section (\ref{eq:4}) is invariant under the interchange
$z \leftrightarrow (1-z)$. Thus we  may replace
the upper limit of the $z$ integration  by $1/2$ and multiply the final
result by 2. Now the upper limit corresponds
to the {\it minimal} value of the pomeron momentum fraction 
$\xpmin$ (or diffractive mass
$M^2=4\,(\pt^2+m_q^2)$) necessary to
produce a dijet system with jets of transverse momenta $\pm\ptbold$. The lower 
limit $\zmin$ corresponds to the {\it maximal} 
allowed value of $\xpmax$ 
(usually $\xpmax < 0.1$).
Thus from (\ref{eq:1}) and (\ref{eq:6}) we obtain
\be
\zmin \;= \; \frac{1}{2}\; 
\biggl\{ 
1-\sqrt{ 1-\frac{4\,(\pt^2+m_q^2)}{\xpmax\,(Q^2+W^2)-Q^2} }\;
\biggr\}\;.
\ee
We now study the  integrations in (\ref{eq:4})
with respect to the azimuthal angles
$\phi$ and $\overline\phi$ between the jet transverse momentum
$\ptbold$ and the gluon momenta $\ktbold$ and $\ktboldbar$, respectively.
The angular structure of the impact factors $\Phi_1$ and
$\Phi_2$ allows us to rewrite the cross section 
formula (\ref{eq:4}) in the following
remarkably compact forms
\bea
\label{eq:13}
\left. \frac{d \sigma_{T}}{d^2 \ptbold  dt}
\right|_{t = 0}  =  \frac{\alpha \alpha_S^2}{3 \pi \ptt} \;
  \sum_q e_q^2
\int_{\zmin}^{1/2} dz \biggl\{ [z^2 + (1 - z)^2 ]~[\phi_1(z,\pt)]^2 
+ \frac{m_q^2}{\pt^2}~[\phi_2(z,\pt)]^2 \biggr\}  
\eea
and
\bea
\label{eq:14}
\left. \frac{d \sigma_{L} }{d^2 \ptbold  dt } 
\right|_{t = 0}  =  \frac{\alpha \alpha_S^2}{3 \pi \ptt} \; 
\sum_q e_q^2
\int_{\zmin}^{1/2} dz  \biggl\{ z^2\, (1-z)^2~\frac{4 Q^2}{\pt^2}~
[\phi_2(z,\pt)]^2 
\biggr\}~,
\\ \nonumber
\eea
where the new ``impact factors'' $\phi_1$ and $\phi_2$  have the following
forms
\be
\label{eq:15}
\phi_1(z,\pt) = \int \frac{d \ktt}{\kt^4}\;
\int\limits_0^{\pi} {d \phi}\; \tilde{f}(x,x^\prime,\ktt)\;
\frac{1}{2}
\biggl\{\frac{1-\omega^2}{1+\omega^2}-\frac{2-a}
{a+b \cos \phi}
\biggr\}\;,
\ee
\be
\label{eq:16}
\phi_2(z,\pt) = \int \frac{d \ktt}{\kt^4}\;
\int\limits_0^{\pi} {d \phi}\; \tilde{f}(x,x^\prime,\ktt)\;
\biggl\{\frac{1}{1+\omega^2}-\frac{1}
{a+b \cos \phi}
\biggr\}~.
\ee
For convenience we have introduced the dimensionless quantities
\be
\label{eq:160}
\omega=\overline{Q}/\,\pt~,~~~~
\tau=\kt/\,\pt~,~~~~
a=1+\omega^2+\tau^2~~~\mbox{\rm and}~~~ b=2\tau~.
\ee 
The cross section formulae (\ref{eq:13}) and (\ref{eq:14}) are quite  general.
They do not depend on any particular behaviour of the gluon distribution
function. 
The crucial observation to obtain these formulae is 
that the terms which could spoil the new impact factor factorization
are odd with respect to the  angles $\phi$ and $\overline\phi$. Thus they give 
a vanishing contribution when the full azimuthal integrations are performed.

\section{Insight from the simplified case}
\label{sec:3}

The angular integrations in (\ref{eq:15}) and (\ref{eq:16}) cannot 
be  performed analytically since the arguments $x$ and $x^\prime$
of  the nonforward
gluon distribution $f$ depend on the angle $\phi$.
It is illuminating, however, to first consider a simplified
situation in which  $\tilde{f}=f(\xp,\ktt)$ depends  on 
$x_{\funp}=x-\xprime$, and not on $x$ and $\xprime$ separately. 
This situation was implicitly assumed in \cite{NZ}.
Then ${f}$ is independent of $\phi$
and the angular integration can be performed, and we obtain
\be
\label{eq:17}
\phi_1(z,\pt) = {\pi}  \;\int \frac{d \ktt}{\kt^4}\;
f(\xp,\ktt)\;
\frac{1}{2}
\biggl\{\frac{1-\omega^2}{1+\omega^2}-\frac{2-a}
{\sqrt{a^2-b^2}}
\biggr\}
\ee
and
\be
\label{eq:18}
\phi_2(z,\pt) = \pi \; \int \frac{d \ktt}{\kt^4}\;
f(\xp,\ktt)\;
\biggl\{\frac{1}{1+\omega^2}-\frac{1}
{\sqrt{a^2-b^2}}
\biggr\}~.
\ee
In the 
limiting case $\omega=\overline{Q}/\,\pt \rightarrow 0$, which corresponds
to diffractive dijet photoproduction with high values of  $\pt$,  
the  only appreciable contribution to the cross section (\ref{eq:13}) 
comes from the $\phi_1$ term. In this limit
$(2-a) \rightarrow (1-\tau^2)$ and $\sqrt(a^2-b^2) \rightarrow |1-\tau^2|$
and so (\ref{eq:17}) becomes simply
\be
\label{eq:18a}
\phi_1(z,\pt) \rightarrow \pi \; \int \frac{d \ktt}{\kt^4}\;
f(\xp,\ktt)\;\Theta(\kt-\pt)~.
\ee
Due to the $1/\,\kt^4$ factor, the dominant contribution to the integral
comes from the region $\kt \gapprox \pt$.
In other words  the  diffractive production of dijets with 
transverse momentum $\pt$ measures  
the gluons forming the QCD pomeron at $\kt \approx \pt$  \cite{NZ}. 
For $\kt<\pt$ the exchanged gluon system is unable to resolve the
$q$ and $\overline{q}$ separately which results in zero net coupling
and is reflected by the theta function in (\ref{eq:18a}). 

\begin{figure}[tb]
   \vspace*{-1cm}
    \centerline{
     \epsfig{figure=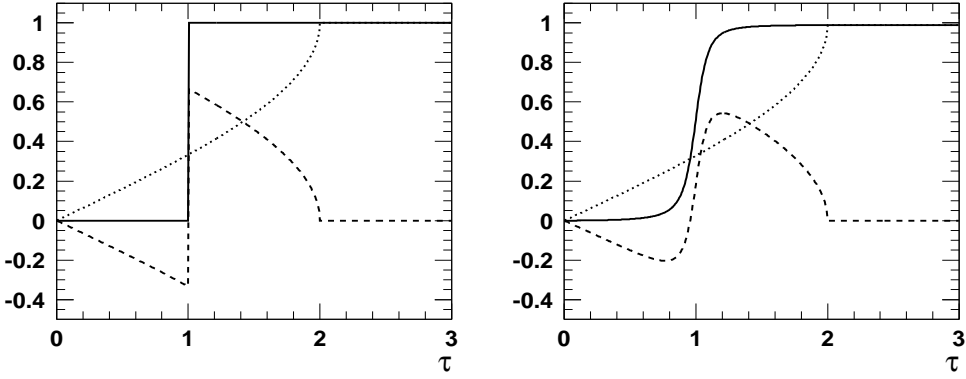,height=7cm,width=15cm}
               }
    \vspace*{-0.5cm}
     \caption{The integrand in the curly brackets 
     in (\ref{eq:17}) as a function of
     $\tau\equiv \kt/\,\pt$ (solid curves). 
     We also show the contributions from the
     $\xprime >0$ (dotted curves) and $\xprime < 0$ (dashed curves) regions
     separately. The two figures correspond 
     to $\omega\equiv\overline{Q}/\,\pt=0$ (left) and $\omega=0.1$ (right).}
\label{fig:1}
\end{figure}

In order to  better understand why the contribution for gluon momenta
$\kt < \pt$ vanishes, we divide the integration in (\ref{eq:15})
into the part corresponding to $\xprime>0$ and the part with $\xprime<0$.
We find from (\ref{eq:7}) that the  condition $\xprime>0$ is fulfilled if  
the azimuthal angle $\phi$ between the jet $\ptbold$ 
and  the gluon $\ktbold$  is given by
\be
\label{eq:19}
0 < \phi < \alpha \equiv \arccos\biggl(-\frac{k_T}{2p_T}\biggr)~,
\ee
whereas the region $\alpha < \phi < \pi$ corresponds to  $x^\prime < 0$.
Notice that if $\kt \ge 2\pt$  then
only the region $x^\prime > 0$  contributes. 
For $\kt < 2\pt$ we split 
the angular integration in (\ref{eq:15}) into the positive and negative
$\xprime$ parts and carry out the integrations separately
\bea
\label{eq:21}
& &\phi_1(z,\pt) = \int \frac{d \ktt}{\kt^4}\;
\biggl\{
\int_0^{\alpha} ...\;+\;\int_{\alpha}^{\pi} ...
\biggl\} =
\frac{\pi}{2} \; \int \frac{d \ktt}{\kt^4}\;
f(\xp,\ktt)\;
\\ \nonumber
\\ \nonumber
&\times& 
\biggl\{ 
\biggl[\frac{1-\omega^2}{1+\omega^2}\biggl(\frac{\alpha}{\pi}\biggr)
-\frac{2-a}{\sqrt{a^2-b^2}}
\biggl(\frac{\alpha \beta}{\pi}\biggr)\biggr]
+
\biggl[\frac{1-\omega^2}{1+\omega^2}\biggl(1-\frac{\alpha}{\pi}\biggr)
-\frac{2-a}{\sqrt{a^2-b^2}}
\biggl(1-\frac{\alpha \beta}{\pi}\biggr)\biggr] 
\biggr\}~,
\eea
where the function $\beta$ is defined by
\be
\label{eq:22}
\beta = \frac{1}{\alpha}
\arccos\biggl({\frac{b+a \cos \alpha}
{a+b \cos \alpha}}\biggr)~.
\ee
Notice that the sum of the two contributions in the squared brackets,
coming from  the $\xprime>0$ and $\xprime<0$ regions, 
simplify in such a way that formula (\ref{eq:17}) is recovered.
In the limiting case when $\omega=0$,   the 
positive and negative $x^\prime$ parts are equal to 
$\alpha\,(1-\beta)/\pi$ and $-\alpha\,(1-\beta)/\pi$ 
respectively, and so their sum gives a vanishing contribution to $\phi_1$
for $\kt < \pt$. As a consequence 
there is no necessity for an infrared cut-off
$k_{T0}$ on the integration in (\ref{eq:17}).
For $\omega \not=0$, however,  the cancellation occurs only
between the contributions from the two $\xprime$ regions that are linear in 
$\kt$. The first nonzero  term
is proportional to $\omega^2\, \kt^2$ and leads to a mild logarithmic
dependence of $\phi_1$ on $k_{T0}$.

The results of the above analysis are well illustrated
by Fig.~\ref{fig:1} where
the solid lines show the integrand in the curly bracket of (\ref{eq:17})
and the dotted and dashed lines show, respectively, 
the $\xprime>0$  and $\xprime<0$ components in (\ref{eq:21}). Both the cancellation
for $\tau < 1$ and the absence of the $\xprime<0$ component for $\kt> 2\pt$
are evident. We take $Q^2=0$ and  $p_T=10~\mbox{\rm GeV}$. 
The two plots correspond to $\omega=\overline{Q}/\,\pt=0$ and 
to $\omega=0.1$ ($m_q=1~\mbox{\rm GeV}$). 

\begin{figure}[tb]
   \vspace*{-1cm}
    \centerline{
     \epsfig{figure=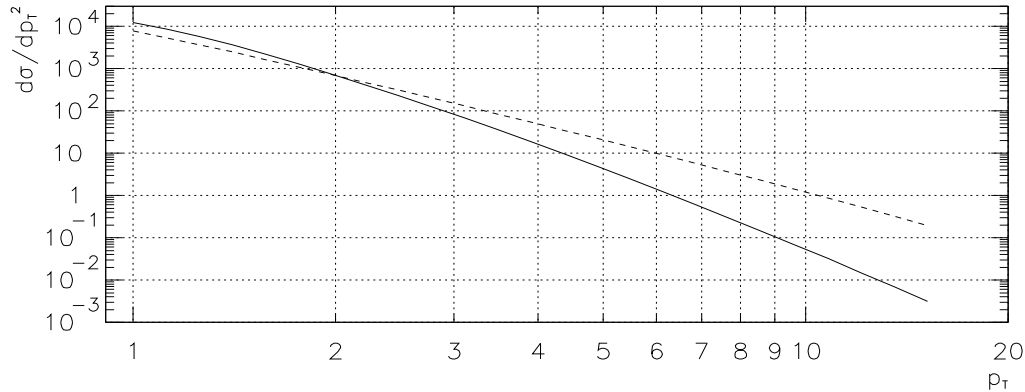,height=7cm,width=15cm}
               }
    \vspace*{-0.5cm}
     \caption{The cross section $d \sigma/d p_T^2$ as a function
     of the jet $p_T$. The continuous curve shows the cross section with
     both the regions $\xprime>0$ and $\xprime<0$ included.
     The dashed curve corresponds
     to the contribution from the  $\xprime > 0$ region only, computed with
     the infrared cut-off $k_{T0}=1 \gev$.
}\label{fig:2}
\end{figure}

It is informative to show the cross section $d\sigma/ d\pt^2$
obtained from (\ref{eq:4}) for the diffractive photoproduction
of dijets in the simplified case when the gluon distribution
$f=1$ and $\alpha_S$ is fixed. The prediction is given by the continuous
curve in Fig.~\ref{fig:2}. At large $\pt$ we anticipate from (\ref{eq:18a}) 
that $\phi_1 \sim 1/\,\pt^2$ and so $d\sigma/ d\pt^2 \sim 1/\,\pt^6$. Such
behaviour is evident in Fig.~\ref{fig:2}. Crucial to this
behaviour is the cancellation between the $\xprime>0$ and $\xprime<0$
contributions that are linear in $\kt$.
This is well illustrated
by the dashed curve in Fig.~\ref{fig:2}, which is the cross section 
which would
have resulted if only the $\xprime>0$ contribution were included.
Then $\phi_1 \sim 1/\,(\pt k_{T0})$ where $k_{T0}$ is the 
infrared cut-off imposed on the $\kt$ integration. 
As a result the cross section behaves as
$1/\,(\pt^4 k_{T0}^2)$ and becomes strongly dependent on the value chosen
for the cut-off  momentum $k_{T0}$. 
For interest we have extended the predictions in Fig.~\ref{fig:2}
down to small values of $\pt$. The large $\pt$ asymptotics 
which we have described
above become apparent for $\pt \gapprox 4~\mbox{\rm GeV}$.

So far our  discussion has been based on the assumption that the gluon
distribution in formulae (\ref{eq:15}) and (\ref{eq:16}) 
does not depend on the azimuthal angle. In this case  
the both $\xprime$ regions are weighted by the same value
of the gluon distribution $f(x_{\funp})$ when the  angular integration
in (\ref{eq:15}) and (\ref{eq:16}) is performed. 
As a result we have the  delicate cancellation
in the region of $\kt < \pt$, which we emphasized above. 
This could be significantly changed when the true off-diagonal 
gluon distribution $f(x,\xprime)$ is used, 
and the two $\xprime$ regions are weighted differently
during the angular integration. The study  of the influence of this 
effect on the diffractive dijet  cross section
is a major objective of our analysis.

\section{Perturbative and nonperturbative contributions}
\label{sec:4}

To evaluate the cross sections  given in (\ref{eq:13}) and 
(\ref{eq:14}) we have to compute the impact factors $\phi_1$ and $\phi_2$
which depend on
the off-diagonal gluon distribution. We divide the $\kt$ integration
in (\ref{eq:15}) and (\ref{eq:16}) into a nonperturbative and perturbative
region according to whether $\kt$ is smaller or greater than 
$k_{T0}\sim 1~\mbox{\rm GeV}$. The angular integration
can be performed analytically in the nonperturbative region. 
For this purpose we analyse the angular dependence of the arguments
$x$, $\xprime$ and  $\xprimetwo$ 
of the gluon distribution function (\ref{eq:16a}) by
dividing  (\ref{eq:7})  by $\xp$ given by (\ref{eq:1}) and (\ref{eq:6})
\be
\label{eq:23}
\frac{\xprime}{\xp} = (1-z)\;
\biggl(\frac{\tau^2 + 2 \tau \cos \phi}{1+\omega^2}
\biggr)\;,
~~~~~~
\frac{x^{\prime\prime}}{\xp} = z\;
\biggl(\frac{\tau^2 + 2 \tau \cos \phi}{1+\omega^2}
\biggr) \;.
\ee 
In the nonperturbative region $\tau=\kt/\pt \ll 1$, and thus we have
\be
\label{eq:24}
|\xprime|,\,|x^{\prime\prime}| \ll \xp
~~~~~~\mbox{\rm and}~~~~~~~x \approx \xp ~.
\ee 
These relations allows us to expand the gluon distribution
(\ref{eq:16a}) around the point $x=\xp$, i.e. we study $\tilde{f}$
as function of $x$ when the value of $\xp$ is fixed.
Thus we obtain
\be
\label{eq:24a}
\tilde{f}(x,\xprime,\kt^2)\;\simeq \;f(\xp,0,\kt^2) \; + \;
\biggl(
\frac{\tau^2/2+\tau \cos \phi}{1+\omega^2}
\biggr)
\;
{\xp} \frac{\partial f}{\partial x}(\xp,0,\kt^2)\;.
\ee
We also expand the expressions in 
the curly brackets in (\ref{eq:15}) and (\ref{eq:16}) 
in powers of $\tau$. Then
the angular integration can be performed analytically 
in the nonperturbative region.
The term linear in  $\tau$  vanishes and 
as a result the first nonzero term is proportional to $\tau^2$ 
(that is $\kt^2$). 
This allows us to express the resulting $\kt$ integration in
terms of the off-diagonal integrated gluon distribution 
at scale $k_{T0}^2$ using 
\bea
\label{eq:27}
\int_0^{k_{T0}^2} \; \frac{d k_T^2}{k_T^2} \; 
{f (\xp, 0, k_T^2)}
\; &=& \; G(\xp,0,k_{T0}^2)~,
\\ \nonumber
\\ 
\int_0^{k_{T0}^2} \; \frac{d k_T^2}{k_T^2} \; 
\frac{\partial f}{\partial x} (\xp, 0, k_T^2)
\; &=& \; \frac{\partial G}{\partial x}(\xp,0,k_{T0}^2)~.
\eea
Note that the off-diagonal gluon distribution
$G(x,0,k^2)$ is not to be identified with the
conventional symmetric gluon distribution $G(x,x,k^2)$
for which the longitudinal fractions are equal 
and hence the asymmetry variable $\xp=0$.

Thus we finally obtain the following decompositions of the impact factors
$\phi_1$ and $\phi_2$ into the following
nonperturbative and perturbative contributions
\bea
\label{eq:28}
\phi_1(z,\pt) &=& 
\frac{\pi}{\pt^2}\;
\biggl\{
\frac{2 \omega^2}{(1+\omega^2)^3}  \;
G(\xp,0, k_{T0}^2) \;+\;
\frac{1-\omega^2}{2 (1+\omega^2)^3}  \;
\xp \frac{\partial G}{\partial x}(\xp,0,k_{T0}^2)
\biggr\}
\nonumber \\
\nonumber \\
\nonumber \\
&+&
\int_{k_{T0}^2}^{\infty} \frac{d \ktt}{\kt^4}\;
\int_0^{\pi} {d \phi}\; \tilde{f}(x,x^\prime,\ktt)\;
\frac{1}{2}
\biggl\{\frac{1-\omega^2}{1+\omega^2}-\frac{2-a}
{a+b \cos \phi}
\biggr\}
\eea
and
\bea
\label{eq:29}
\phi_2(z,\pt) &=&\frac{\pi}{\pt^2}  \;
\biggl\{
\frac{\omega^2-1}{(1+\omega^2)^3}~G(\xp,0, k_{T0}^2)  \;+\;
\frac{1}{(1+\omega^2)^3} \;
\xp \frac{\partial G}{\partial x}(\xp,0,k_{T0}^2)
\biggr\}
\nonumber \\
\nonumber \\
\nonumber \\
&+&
\int_{k_{T0}^2}^{\infty} \frac{d \ktt}{\kt^4}\;
\int_0^{\pi} {d \phi}\; \tilde{f}(x,x^\prime,\ktt)\;
\biggl\{\frac{1}{1+\omega^2}-\frac{1}
{a+b \cos \phi}
\biggr\}~.
\eea
For diffractive dijet production with large $\pt$ by far the dominant
contribution comes from the perturbative region.

Relations (\ref{eq:24}) are not generally valid in the perturbative region. 
For example 
for gluon transverse momenta  $\kt \approx \pt$, which dominate
in the simplified case of Section \ref{sec:3},  
we find that at least one of the fractions
$|\xprime|,\,|\xprimetwo| \sim \xp$. For  $z=1/2$ and 
$\omega \ll 1$ this is equivalent to
\be
\label{eq:29a}
1/2\;\xp\;\lapprox\;x\;\lapprox\;5/2\;\xp\;.
\ee
Thus in the perturbative region we study a much broader range of $x$ 
of the off-diagonal gluon distributions (\ref{eq:16a}) 
with the second gluon momentum
fractions $\xprime$ or  $\xprimetwo$ being both positive
and negative. Also
the off-diagonal distributions are required over quite a range of $\kt$
and are computed using evolution equations. We discuss the off-diagonal
evolution in detail in the next section.

\section{Off-diagonal parton distributions and their evo\-lu\-tion}
\label{sec:5}

Before discussing the evolution equation for the off-diagonal
parton distribution it is convenient to follow Radyushkin \cite{RAD1,RAD2}
and use the notation\footnote{Alternative formulations of
the off-diagonal evolution equations can be found 
in \cite{FS} and \cite{JI2}.}
\be
\label{eq:31}
G_{\xp}(x,\kt) \equiv G(x,\xprime,\kt^2)~,
\ee
for the off-diagonal gluon 
distribution,  where the dependence 
on $\xprime$ is implicit through the relation $\xprime=x-\xp$.
Both the variables $x$ and $\xp$ lie in the interval $0$ to $1$.
Therefore we view the off-diagonal  distributions as 
a  family of functions (of $x$ and the scale $\kt$)
labelled by the asymmetry variable $\xp$.
Radyushkin uses $\zeta$ for $\xp$,
but for the moment we will keep the latter notation to indicate that 
$\xp$ defined in (\ref{eq:1}) is the asymmetry variable 
in the process that we consider.

In the region of $x>\xp$ (that is $\xprime>0$)
the off-diagonal gluon distribution $G_{\xp} (x)$ describes
a gluon emitted by the proton with  momentum fraction $x$ together with
a gluon absorbed with fraction $\xprime=x-\xp$. Thus 
$G_{\xp} (x)$ is a generalization of the ordinary diagonal 
gluon distribution,  that is in the diagonal limit 
$\xp\equiv x-\xprime \rightarrow 0$ we have
\be
\label{eq:32}
G_{\xp=0} (x) =  G(x,x) \equiv x\, g(x)~.
\ee
We stress once again that we must distinguish between
$G(x,x)$ and $G_{x}(x)\equiv G(x,0)$. In the first case the asymmetry variable
$\xp=0$ while in the later $\xp=x$.

In the region  $x<\xp$ the second gluon has a {\it negative} fraction 
$\xprime$ of the proton's momentum. In this case it is more transparent
to use $-k^{\prime}$ for the momentum
of the second gluon, that is we make the replacement
\be
\label{eq:33}
k^\prime = \xprime\, p + \kt~~~\rightarrow~~~~
\tilde{k}^\prime=(-k^\prime)= |\xprime|\, p - \kt~. 
\ee
Thus we have the two gluons emitted form the proton with the momenta
$k=x\, p + \kt$ and $\tilde{k}^\prime$.
The gluon distribution $G_{\xp}(x)$ for $x < \xp$ 
may be regarded as the probability amplitude 
for emitting a two gluon colour singlet system 
from the proton with the individual gluons 
carrying fractions $x/\,\xp$ and $|\xprime|/\,\xp$ of the total momentum
of the system $r=\xp\, p=k+\tilde{k}^\prime$.

To evaluate the perturbative contribution we use the full
form of the off-diagonal
evolution equations for the gluon distribution $G_{\xp}(x,\mu)$
($\mu=\kt$ in our case) and the off-diagonal singlet quark distribution
\be
\label{eq:35}
\Sigma_{\xp}(x,\mu)=\sum_{i=1}^{N_f}\;\bigl\{q_{\xp}^i (x,\mu)
+ \overline{q}_{\xp}^i (x,\mu) \bigr\}~,
\ee
where $q_{\xp}^i$ and $\overline{q}_{\xp}^i$ are off-diagonal
quark and antiquark distributions introduced in analogy to the diagonal case.
The evolution equations have the following general form
\bea
\label{eq:36}
\mu \frac{\partial}{\partial \mu}\;\Sigma_{\xp}(x,\mu) =
\int\limits_0^1 dz\; P^{QQ}_{\xp}(x,z;\mu)\;\Sigma_{\xp}(z,\mu)  +
\int\limits_0^1 dz\; P^{QG}_{\xp}(x,z;\mu)\; G_{\xp}(z,\mu)
\nonumber \\
 \\
\mu\frac{\partial}{\partial \mu}\;G_{\xp}(x,\mu) =
\int\limits_0^1 dz\; P^{GQ}_{\xp}(x,z;\mu)\;\Sigma_{\xp}(z,\mu)  +
\int\limits_0^1 dz\; P^{GG}_{\xp}(x,z;\mu)\;G_{\xp}(z,\mu) ,
\nonumber
\eea
where the structure of the kernels $P$ depends on the relations between
the $x,z$ and $\xp$ variables. The detailed form of the evolution equations
extracted from  the review \cite{RAD2} is given in  the Appendix.

For $x>\xp$ ($\xprime>0$) the integration in (\ref{eq:36}) covers the range
from $x$ to $1$ and the kernels have a form which
coincides with the standard Altarelli-Parisi  kernels
for $\xp=0$.
That is (\ref{eq:36}) reduce to the 
DGLAP evolution equations \cite{DGLAP} in the limit $\xp=0$. On the other hand
for $x<\xp$ ($\xprime<0$) the form of the kernels
depends on whether the integration variable $z$ is less or greater
than $x$. In this case the integration covers  the whole range $(0,1)$
and in the the limit $\xp=1$ Eqs.~(\ref{eq:36}) reduce
to the ERBL evolution equations \cite{ER,BL} for 
the parton distribution amplitudes in a meson.

We solve the evolution equations numerically using an expansion in
terms of Chebyshev polynomials and cross-check the solutions
against, on the one hand,  
the known asymptotic solutions of the ERBL equations
and, on the other hand,  
the solutions of the DGLAP evolution equations. Moreover  particular
attention is paid to the behaviour of the resulting 
off-diagonal parton distributions at point $x=\xp$, where the two 
(DGLAP-like and ERBL-like) forms of the off-diagonal
evolution equations coincide. In all cases we find  
excellent agreement with
the expectations resulting from analytical insight. 

\begin{figure}[t]
   \vspace*{-1cm}
    \centerline{
     \epsfig{figure=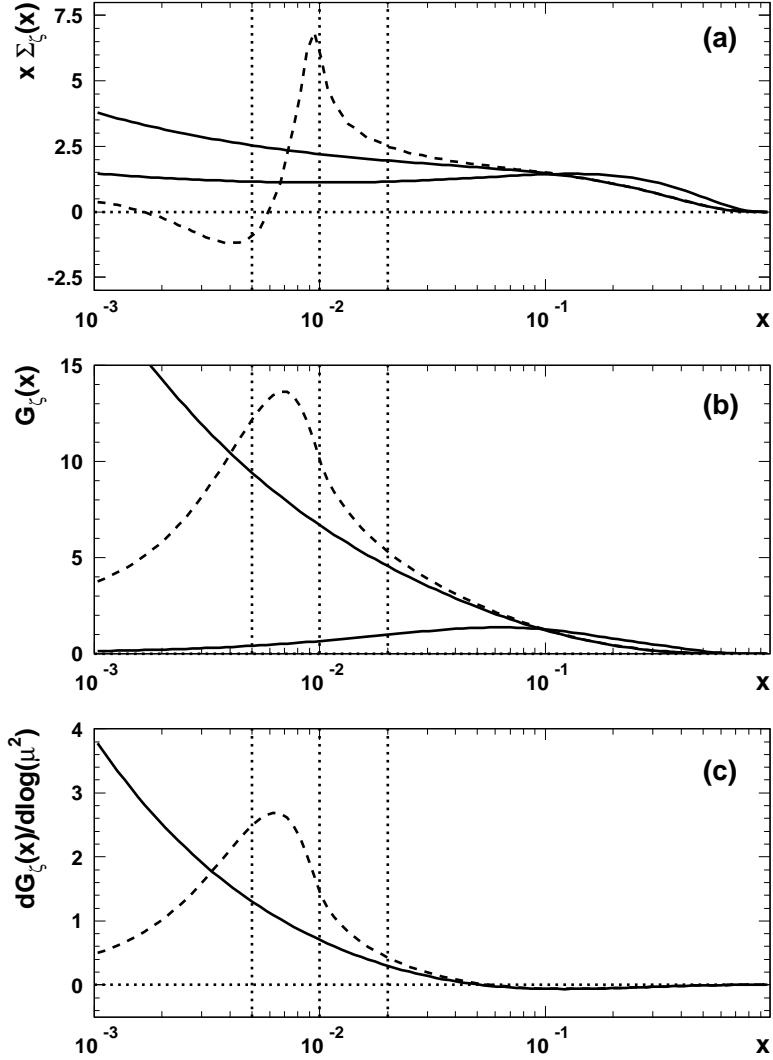,height=15cm,width=13cm}
               }
    \vspace*{-0.5cm}
     \caption{The off-diagonal      
     singlet and gluon distributions and
      the logarithmic derivative of the gluon distribution
     (dashed curves) at $\mu^2=10^2~\mbox{\rm GeV}^2$  for 
     $\zeta\equiv\xp=10^{-2}$ (middle vertical
     line). The continuous curves show
     the initial distributions at $\mu_0^2=1~\mbox{\rm GeV}^2$
     (lower curves) 
     and effect of their evolution using the DGLAP equations
     (upper curves).    
}
\label{fig:3}
\end{figure}

In the two upper plots of 
Fig.~\ref{fig:3} 
we show the off-diagonal singlet and gluon  distributions
evolved up to $\mu^2=100~\mbox{\rm GeV}^2$ (dashed lines)
starting from initial conditions specified
at $\mu_0^2=1~\mbox{\rm GeV}^2$ (lower solid curves).
We  postulate the following form of the initial distributions
\bea
\nonumber
\label{eq:37}
\Sigma_{\xp}(x) &=& (1-\xp)^n\; \bigl\{\;(1-\xp)\; \Sigma_{\mrs}(x) +
                             \xp \; \Sigma_{\bl}(x)\;\bigr\}\;,
\\ 
\\ \nonumber
G_{\xp}(x) &=& (1-\xp)^m\;\bigl\{\;(1-\xp)\; G_{\mrs}(x) +
                             \xp \; G_{\bl}(x)\;\bigr\}\;,
\eea
where $\Sigma_{\mrs}$ and $G_{\mrs}$ are the MRS distributions \cite{MRS}
obtained from  global fits based on the DGLAP evolution equations
to data for deep inelastic and related hard scattering processes, and 
$\Sigma_{\bl} \sim x\, (1-x)$ and $G_{\bl} \sim x^2\, (1-x)^2$ are 
asymptotic solutions of the ERBL equations (for the nonsinglet
and pure gluon cases). 
The postulated form in the curly brackets 
encompasses the basic feature of the off-diagonal parton distributions
(and evolution equations) that in the limits 
$\xp \rightarrow 0~\mbox{\rm or}~1$ the 
DGLAP or ERBL components, respectively,  are obtained. 
However in case of a nucleon
it is unlikely that the two partons can be {\it emitted}, 
sharing the whole nucleon longitudinal momentum 
(which corresponds to the $\xp=1$ limit).
This observation is accounted
for by the additional powers of $(1-\xp)$ with $n,m>0$.
The asymmetry variable was chosen to be
$\xp=10^{-2}$, and is indicated by the middle vertical dotted
line (the outermost vertical lines correspond to the values
$\xp/\,2$ and $2\,\xp$). Clearly the initial distributions are dominated
by the DGLAP (MRS) component for such a small value of $\xp$.
We may compare the dashed curves 
(the evolved off-diagonal parton distributions)
with the upper continuous curves
which are obtained from the same initial conditions using
the conventional DGLAP evolution equations\footnote{The off-diagonal 
gluon distribution (Fig.~\ref{fig:3}b) in the
$\xprime>0$ region, to the right of the central vertical line, 
together with the
conventional DGLAP curve reproduce the main features of the 
earlier study of Ref.~\cite{MR}.}.
The bottom plot compares the logarithmic derivative of $G_{\xp}(x)$,
which is used in our analysis, see (\ref{eq:3}),  
at the same scale $\mu^2=100~\mbox{\rm GeV}^2$. Again the dashed 
curve corresponds to 
the off-diagonal case while the continuous curve is obtained 
using the DGLAP evolution equations.

The solutions reflect the mixed (DGLAP and ERBL) nature
of the off-diagonal evolution.
This is particularly visible for $\Sigma_{\zeta=\xp}(x)$
which becomes {\it negative} for $x<\xp$. This feature is found
for the solution of the ERBL evolution equations for 
the distribution amplitude
which is obtained from $\Sigma_{\zeta}(x)$ in the limit $\zeta=\xp=1$.
Thus  the off-diagonal parton distributions do not have a probabilistic
interpretation, unlike the conventional diagonal distributions. The region 
$|x-\xp| \sim \xp$, indicated by the three vertical lines, 
is particularly interesting from the point of view
of the difference between the off-diagonal and diagonal parton distributions.
This region is relevant for  gluons with momenta $\kt\approx \pt$ 
(see (\ref{eq:29a}))
which are supposed to give an important contribution to the 
dijet production cross section.
We will explore this effect in the next section.  

\section{Effect of off-diagonal distributions on the dijet cross section}
\label{sec:6}

\begin{figure}[t]
   \vspace*{-1cm}
    \centerline{
     \epsfig{figure=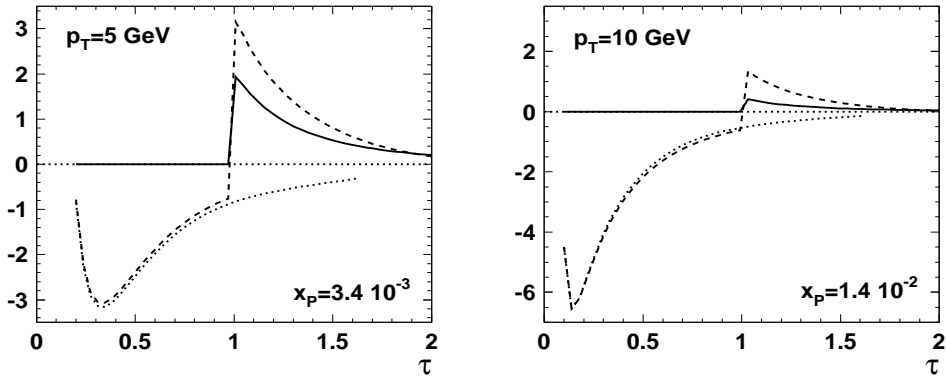,height=7cm,width=15cm}
               }
    \vspace*{-0.5cm}
     \caption{The integrand in the square brackets of (\ref{eq:38})
     for two values of dijet momentum $\pt$
     and the corresponding minimal values of $\xp$. The continuous
     curves show the integrand in the simplified case when
     $\tilde{f}_{\xp}=\tilde{f}_{\xp}(\xp)$, while the dashed curves
     show the full off-diagonal case  $\tilde{f}_{\xp}=\tilde{f}_{\xp}(x)$.
     The dotted curves adjacent to  the dashed ones
     show the approximate integrand of (\ref{eq:39}) when only the term 
     linear in $\tau$ is retained.
     }
\label{fig:4}
\end{figure}

For diffractive dijet 
photoproduction only the cross section (\ref{eq:13}) for the
transversely polarised photons contributes. In addition the $\phi_2$
part is strongly suppressed for high values of $\pt$. 
Thus we need only analyse
the impact factor $\phi_1$ given by (\ref{eq:28}).
In terms of $\tau=\kt/\pt$ the perturbative part of $\phi_1$ 
has  the following form for $\omega=0$
\be
\label{eq:38}
\phi_{1}^{\mbox{\scriptsize (pert)}}(z,\pt)\;=\;
\frac{2\pi}{\pt^2}\;
\int_{\tau_0}^{\infty} d\tau\;
\biggl[
\frac{1}{\tau^3}
\int_0^{\pi} \frac{d \phi}{\pi}\; \tilde{f}_{\xp}(x,\tau \pt)\;
\frac{1}{2}
\biggl\{1\;-\;\frac{1-\tau^2}{1+\tau^2+2\tau \cos \phi}
\biggr\}
\biggr]\;.
\ee
where $\tau_0=k_{T0}/\pt$.
The function $\tilde{f}_{\xp}$ is written in the notation
introduced by Eq.~(\ref{eq:31}) and is related to the off-diagonal
gluon distribution $G_{\xp}$ through 
relations (\ref{eq:3}) and (\ref{eq:16a}).
In Fig.~{\ref{fig:4}} we show the integrand in the squared brackets in
(\ref{eq:38}) as a function of $\tau$.
We choose for illustration $\pt=5~\mbox{\rm and}~10\gev$ for
dijet transverse momentum. 
For each choice we assume
the minimum value of $\xp=4\pt^2/\,W^2$, which corresponds
to  $z=1/2$.

The results depend subtlely on the properties 
of the gluon distribution $\tilde{f}_{\xp}$. 
Let us start with the simplified case discussed
in Section \ref{sec:3}. That is we assume that 
$\tilde{f}_{\xp}={f}_{\xp}(\xp,\tau\pt)$ in (\ref{eq:38}), hence
the gluon distribution does not depend on the azimuthal angle. DGLAP
evolution  is appropriate in this case and the resulting 
integrand in  (\ref{eq:38}) is shown by the continuous 
curves in Fig.~{\ref{fig:4}}.
The angular integration leads to the step-like form of the integrand
which is then modified by the $1/\tau^3$ factor 
to give a function peaked at $\tau \approx 1$. This result was
anticipated in Section \ref{sec:3}.

Now let us explore what happens to the integrand when we use the full
off-diagonal gluon distribution $\tilde{f}_{\xp}(x,\tau\pt)$,  evolved
using Eqs.(\ref{eq:36}) from the input given in (\ref{eq:37}). The
results are shown by the dashed curves and are strikingly different
from the continuous curves of the simplified case. First we see
that for $\tau \gapprox 1$ the integrand is larger. This can be
easily  understood by inspecting the off-diagonal gluon distribution
shown in Fig.~{\ref{fig:3}}c. If $\tau>1$ (that is $\kt>\pt$) then during
the angular integration the argument $x$ of the gluon $\tilde{f}_{\xp}(x)$
spans approximately the region indicated by the outer vertical lines
with the gluon following the dashed curves, see also (\ref{eq:29a}).
In contrast
in the simplified case, the function $\tilde{f}_{\xp}(\xp)$ is constant
as $x$ spans  the allowed range, with its value given by where
the solid curve crosses the central vertical line 
in Fig.~{\ref{fig:3}}c.
The difference in these forms of $\tilde{f}_{\xp}$ explains why
the dashed curve in Fig.~{\ref{fig:4}} is larger than the
continuous curve for $\tau \gapprox 1$. 

The biggest change is in the region $\tau<1$.
Recall that in the simplified case the regions 
of  $\xprime >0$ and $\xprime < 0$
(to the right and to the left, respectively, of the central 
vertical line in Fig.~\ref{fig:3}c)
are weighted by the same value ${f}_{\xp}(\xp)$ when
the angular integration is performed. As a result the terms linear
in $\tau$ coming from the two $\xprime$ regions cancel 
(see Fig.~\ref{fig:1} for
illustration of this mechanism). However
the off-diagonal distributions
are asymmetric with respect to the line $x=\xp$ which
divides the two $\xprime$ regions (see dashed curve in Fig.~\ref{fig:3}c). 
This leads the lack of the above mentioned cancellation and in consequence
to an important contribution 
for $\tau <1$, shown by the dashed lines in Fig.~\ref{fig:4}. 
We quantify this observation by expanding the function in the curly brackets
in (\ref{eq:38}) in powers of $\tau$
\be
\label{eq:39}
\phi_{1}^{\mbox{\scriptsize (pert)}}(\tau<1)\; \simeq\;
\frac{2\pi}{\pt^2}\;
\int\limits_{\tau_0}^1 d\tau\;
\biggl[
\frac{1}{\tau^3}
\int\limits_{0}^{\pi} \frac{d\phi}{\pi}\; 
\tilde{f}_{\xp}(x,\tau\pt)\;
\bigl\{
\tau \cos \phi - \tau^2 \cos 2\phi\; +\; ...\;
\bigr\}
\biggr]\;.
\ee
Now we retain the term linear in $\tau$ and perform the angular integration.
The resulting integrands are the dotted lines in Fig.~\ref{fig:4},
extended into the $\tau > 1$ region to make them visible. 
Thus in the off-diagonal case the linear
in $\tau$ term from the above expansion does not vanish after the angular
integration as it did in the simplified case. 
Moreover the whole contribution from  the $\tau<1$ region is essentially 
determined by this term.

\begin{figure}[t]
   \vspace*{-1cm}
    \centerline{
     \epsfig{figure=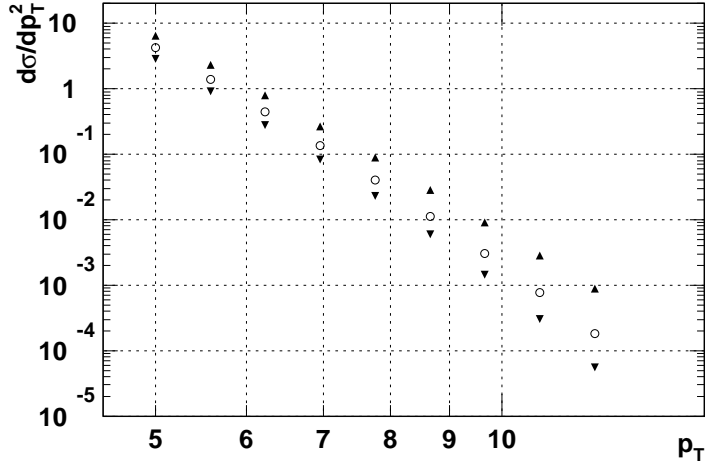,height=8cm,width=12cm}
               }
    \vspace*{-0.5cm}
     \caption{The diffractive dijet photoproduction cross section
     (in $\mbox{\rm pb/GeV}^2$) integrated over the interval
     $\xpmin<\xp<0.05$
     as a function of the jet transverse momentum (in GeV). 
     The upper triangular points are obtained in the off-diagonal
     analysis and the lower ones correspond to the simplified case. 
     The open points are obtained for  
     $\tilde{f}_{\xp}(x)$ evolved with the DGLAP equations.
     }
\label{fig:5}
\end{figure}

At this point we may worry that the contribution coming from the linear
in $\tau$ term in (\ref{eq:39}) 
drastically increases the sensitivity 
of  $\phi_{1}^{\mbox{\scriptsize (pert)}}$,
and in consequence the dijet cross  section,    
to the choice of the infrared parameter $k_{T0}$. 
Fortunately this does not happen. The reason is as follows. For
$\tau_0 < \tau \ll 1$  we may additionally expand the 
function $\tilde{f}_{\xp}(x)$
in (\ref{eq:39}) around $x=\xp$ using formula (\ref{eq:24a}). Then we obtain
\bea
\nonumber
\label{eq:40}
\phi_{1}^{\mbox{\scriptsize (pert)}}(\tau \ll 1) &\simeq&
\frac{2\pi}{\pt^2}\;
\int\limits_{\tau_0} \frac{d\tau}{\tau^3}
\int\limits_{0}^{\pi} \frac{d\phi}{\pi} 
\biggl\{
{f}_{\xp}(\xp,\tau\pt)\; +\; 
\tau \cos \phi\;
\left.\frac{\partial {f}_{\xp}(x,\tau\pt)}{\partial \ln x}
\right|_{x=\xp}
\biggr\}
\biggl\{
\tau \cos \phi
\biggr\}
\\ \nonumber
\\
&=& \frac{2\pi}{\pt^2}\; \int_{\tau_0} {d\tau}\;\frac{1}{2\tau}\;
\left. \frac{\partial 
{f}_{\xp}(x,\tau\pt)}{\partial \ln x}\right|_{x=\xp}\;,
\eea
where the angular integration has been performed 
to give the final result. Notice
that in the simplified case of Section \ref{sec:3} 
only the first term in the expansion
of $\tilde{f}_{\xp}$ exists and the vanishing contribution is obtained
after the angular integration. The  full  off-diagonal treatment
introduces an additional linear in $\tau$ term 
which leads to the first nonvanishing contribution being 
quadratic in $\tau$. As a result
$\phi_1^{\mbox{\scriptsize (pert)}}$ still depends 
at most logarithmically on the
parameter $k_{T0}$. We have checked that the integrand in (\ref{eq:40})
reproduces to a good approximation  the exact integrand in (\ref{eq:38}) 
up to $\tau \approx 0.3$ for both the values of $\pt$ chosen for 
Fig.~\ref{fig:4}. 

The above analysis also shows that the size of the negative
contribution to $\phi_{1}^{\mbox{\scriptsize (pert)}}$ 
is mostly determined by the value of logarithmic derivative in $x$ of
$f_{\xp}(x)$, taken at the point $x=\xp$. In Fig.~\ref{fig:3}c
this derivative corresponds to the slope of 
the dashed curve at the intersection point
with the middle vertical line. On inspecting Fig.~\ref{fig:3}c notice that
the gluon distribution  evolved with the DGLAP equations
(continuous curve) is also asymmetric with respect to the line $x=\xp$. Thus,  
when such gluon is used in the dijet analysis, 
a similar negative contribution will be 
present. The size of this contribution, however, is
 smaller than in the fully off-diagonal
case since the slope of the continuous 
curve at $x=\xp$ is always smaller than that of
the dashed (off-diagonal) one.

In summary, 
the off-diagonal parton distributions enhance the contribution
to $\phi_1$ in comparison to the simplified case for $\kt \ge \pt$
and  lead to a significant {\it negative} contribution in the region
$\kt < \pt$. Both contributions reflect the particular form of
the gluon distribution (\ref{eq:3}) that is dictated by 
the off-diagonal evolution equations (\ref{eq:36}). We have checked
that for $\pt \ge 5\gev$ the net effect after the $\tau$ ($\kt$) integration
in (\ref{eq:38}) is always negative in the whole range of the $z$ variable
($\zmin<z<1/2$) in the fully off-diagonal case.  Moreover, after squaring
and integrating over $z$ in (\ref{eq:13}),
the impact factor $\phi_1$ computed using 
the true off-diagonal distributions leads to 
a significantly larger cross section 
for diffractive dijet photoproduction (\ref{eq:13}) 
than in the simplified case.
In Fig.~\ref{fig:5} we show the cross section $d\sigma/d\pt^2$,
integrated over  $\xp$  between the values $\xpmin=4\pt^2/\,W^2$ and
$\xpmax=0.05$. In addition we 
approximate the effect of the integration over the momentum transfer $t$
by dividing (\ref{eq:13}) by the diffractive slope $A_D=6 \gev^{-2}$.
The upper triangular points correspond to the true off-diagonal result while
the lower ones are obtained in the simplified analysis. The open points
show the result when $\tilde{f}_{\xp}(x,\kt)$ is evolved  using 
the DGLAP equations. 

We emphasize the special nature of the exclusive diffractive photoproduction
of dijets as a probe of the off-diagonal gluon distribution. In our study
we have assumed that the cross section is dominated  by the QCD
subprocess $\gamma p \rightarrow (q\overline{q})\, p^{\prime}$. 
A possible contamination may arise 
from the subprocess $\gamma p \rightarrow (q\overline{q}g)\, p^{\prime}$,
in which the $q\overline{q}$ pair and the gluon  form the two jet system.
This background process has been estimated in \cite{WUST} for 
the inclusive diffractive
cross section and found to be important in the region of low $\pt$ and large
diffractive mass $M$, 
but its contribution should be small in our case since we are
studying  the large $\pt$ domain.

\section{Conclusions}
\label{sec:7}

The diffractive photoproduction of dijets with high $\pt$ is an ideal probe
of the properties of the off-diagonal (unintegrated) gluon distribution 
$f(x,\xprime,\kt^2)$ 
over a wide kinematic range. Indeed the calculation of the differential 
cross section 
$d\sigma/d\pt^2$ involves the integration over entire range of $\kt$ with important
contributions coming from the region $\kt \gapprox \pt$, and so requires 
the solutions
of the off-diagonal evolution equations to determine $f$. Our detailed studies of
these equations revealed some novel features. We found that the region $\kt<\pt$ 
gives an important contribution to the cross section, originating from the specific
properties of the evolved gluon in the $\xprime<0$ (ERBL) and $\xprime>0$ (DGLAP)
domains. This is in contrast to
the simplified treatment, discussed in Section \ref{sec:3}, in which the entire
contribution to the cross section
comes  from the region $\kt \gapprox \pt$ leading to a far smaller cross section. 
In the  simplified analysis 
the gluon distribution was assumed to be only a function of $\xp$ (and $\kt^2$) --
an equivalent approximation was made in the analysis of Ref.~\cite{NZ}.

To summarize we found that if the true off-diagonal gluon distribution is used
then the impact factor $\phi_1$ (Eq.~(\ref{eq:28})) 
receives an important negative contribution from the $\kt<\pt$ region as well as an
enhanced positive contribution from the $\kt>\pt$ domain. 
The negative contribution increasingly dominates with growing $\pt$.
After integration over
$\kt$ the net effect is an overall enhancement of the dijet cross section
in comparison with either
the simplified case or with the case in which the conventional
DGLAP evolution is used. For example, the enhancement of $d\sigma/d\pt^2$ at 
$\pt=10\gev$ due to off-diagonal effects is a factor of $3$ as compared to using
DGLAP evolution, and a factor of $6$ compared to the simplified treatment of
Section \ref{sec:3} and Ref.~\cite{NZ}. The corresponding factors for jets of
$\pt=5\gev$ are $1.5$ and $2.2$ respectively. These enhancements shown in Fig.~\ref{fig:5}.
can, in principle, be tested experimentally offering an ideal testing ground of
off-diagonal effects.

\vskip 1cm
\centerline{\large \bf Acknowledgements}

We thank A. V. Radyushkin, 
M. G. Ryskin and M. W\"{u}sthoff for valuable discussions.  
KGB and JK
thank, respectively the Royal Society/NATO and the UK Particle Physics and 
Astronomy Research Council for Fellowships. They also thank the Department
of Physics of the University of Durham  and Grey College 
for their warm hospitality.
This research has been supported
in part the Polish State Committee for Scientific Research grant no. 2 P03B 089 13.

\newpage
\section*{Appendix}

Here we present for  reference the full form of the evolution equations
for the off-diagonal singlet and gluon distributions $\Sigma_{\zeta}(x,\mu)$
and $G_{\zeta}(x,\mu)$, following the prescriptions for the kernels given
by Radyushkin in \cite{RAD2}. We use his definitions and notations 
of off-diagonal (asymmetric in his language) parton distributions.
In this notation $\zeta$ is the asymmetry (skewedness) parameter 
which in the main body of the paper is equal to the variable $\xp$
\be
\label{eq:a0}
\zeta\equiv x-\xprime=\xp~,
\ee
with $0 \le \zeta \le 1$.
The distributions are defined in such way that in the
limit $\zeta\rightarrow 0$ (the DGLAP limit) we obtain the ordinary
{\it parton distributions}
\bea
\label{eq:a1}
\Sigma_{\zeta}(x,\mu)~&\rightarrow&~\sum_{i=1}^{N_f}~\bigl\{
q^i(x,\mu)+\overline{q}^i(x,\mu)
\bigr\}
\\
\nonumber \\
G_{\zeta}(x,\mu)~&\rightarrow&~x~g(x,\mu)~.
\eea
Notice that the ordinary quark distributions are not multiplied by $x$
as in the gluon case. In the opposite limit $\zeta\rightarrow 1$ 
(the ERBL limit)
the off-diagonal distributions become the {\it distribution  amplitudes}
for finding two partons in a meson sharing fractions $x$ and $1-x$
of its momentum.
\bea
\label{eq:a2}
\Sigma_{\zeta}(x,\mu)~&\rightarrow&~\sum_{i=1}^{N_f}~\bigl\{
\Psi^i(x,\mu)+\overline{\Psi}^i(x,\mu)
\bigr\}
\\
\nonumber \\
G_{\zeta}(x,\mu)~&\rightarrow&~\Psi^g(x,\mu)~.
\eea
Using the notation $\xprime=x-\zeta$ and $\zprime=z-\zeta$
we have the following equations for $x>\zeta$ 
(that is for $\xprime >0$)
\bea
\label{eq:a3}
\mu\frac{\partial}{\partial \mu}\Sigma_{\zeta}(x,\mu)
&=&
\frac{\alpha_S(\mu)}{\pi}  C_F\;
\biggl\{
\int\limits_x^1 \frac{dz}{x-z}\;
\biggl[
\biggl(\frac{x}{z}+\frac{\xprime}{\zprime}\biggr)\;\Sigma_{\zeta}(x,\mu)-
\biggl(1+\frac{x \xprime}{z \zprime}\biggr)\;\Sigma_{\zeta}(z,\mu)
\biggr]
\nonumber \\
\nonumber \\
& &\qquad\qquad\quad +\;
\Sigma_{\zeta}(x,\mu)\;\biggl[\;\frac{3}{2} + \ln\frac{(1-x)^2}{1-\zeta}
\biggr]
\biggr\}
\nonumber \\
\nonumber \\
&+&
\frac{\alpha_S(\mu)}{\pi} N_f\; 
\int\limits_x^1 \frac{dz}{z \zprime}\;
\biggl[\biggl(1-\frac{x}{z}\biggr)\biggl(1-\frac{\xprime}{\zprime}\biggr)\;
+\frac{x \xprime}{z \zprime}
\;\biggr]\;G_{\zeta}(z,\mu)\;,
\eea


\bea
\label{eq:a4}
\mu\frac{\partial}{\partial \mu}G_{\zeta}(x,\mu)
&=&
\frac{\alpha_S(\mu)}{\pi} C_F\;
\int\limits_x^1 dz\; \biggl[\biggl(1-\frac{x}{z}\biggr)
\biggl(1-\frac{\xprime}{\zprime}\biggr) +1
\biggr]\;\Sigma_{\zeta}(z,\mu)
\nonumber \\
\nonumber \\
&+&
\frac{\alpha_S(\mu)}{\pi} N_c\;
\biggl\{
\int\limits_x^1 dz\; 
\biggl[
\frac{2}{z} \biggl(1+\frac{x \xprime}{z \zprime}\biggr)
\biggl(1-\frac{\xprime}{\zprime}\biggr)\;G_{\zeta}(z,\mu)
\nonumber \\
\nonumber \\ 
&+&
\frac{\bigl[({x}/{z})+({\xprime}/{\zprime})\bigr]\; G_{\zeta}(x,\mu)
-\bigl[({x}/{z})^2+({\xprime}/{\zprime})^2\bigr]\; G_{\zeta}(z,\mu)}
{x-z}
\biggr]
\nonumber \\
\nonumber \\
& &\qquad\qquad\quad +\;
G_{\zeta}(x,\mu)\;
\biggl[\frac{11-2/(3N_f)}{2N_c}+\ln\frac{(1-x)^2}{1-\zeta}
\biggr]
\biggr\}~,
\eea
where $C_F=4/3$ and $N_c=3$, and $N_f$ is the number of active flavours.
In the limit $\zeta=0$ the above equations become the DGLAP evolution
equations.


The equations for $x<\zeta$ (that is for $\xprime<0$)
are more complicated since they
involve integration with different kernels in the intervals
$(0,x)$ and $(x,1)$. We have 
\bea
\label{eq:a5}
\mu\frac{\partial}{\partial \mu}\Sigma_{\zeta}(x,\mu)
&=&
\frac{\alpha_S(\mu)}{\pi}  C_F\;
\biggl\{
\int\limits_0^x dz\; \biggl(\frac{\xprime}{\zprime} \biggr)
\;\biggl[
\frac{\Sigma_{\zeta}(z,\mu)}{\zeta}+
\frac{\Sigma_{\zeta}(z,\mu)-\Sigma_{\zeta}(x,\mu)}{x-z}
\biggr]
\nonumber \\
\nonumber \\
& &\qquad\qquad\quad +\;
\int\limits_x^1 dz\; \biggl( \frac{x}{z}\biggr)
\biggl[
\frac{\Sigma_{\zeta}(z,\mu)}{\zeta}+
\frac{\Sigma_{\zeta}(z,\mu)-\Sigma_{\zeta}(x,\mu)}{z-x}
\biggr]
\nonumber \\
\nonumber \\
& &\qquad\qquad\quad +\;
\Sigma_{\zeta}(x,\mu)\;
\biggl[
\frac{3}{2}+\ln\frac{x(1-x)}{\zeta}
\biggr]
\biggr\}
\nonumber \\
\nonumber \\
&+& 
\frac{\alpha_S(\mu)}{\pi} N_f\;
\biggl\{
\int\limits_0^x \frac{dz}{\zeta^2}\; 
\biggl(\frac{\xprime}{\zprime} \biggr)
\biggl[
\;4 \frac{x}{\zeta} + \frac{2x-\zeta}{\zeta-z}
\biggr]\;G_{\zeta}(z,\mu)
\nonumber \\
\nonumber \\ 
& &\qquad\qquad\quad -\;
\int\limits_x^1 \frac{dz}{\zeta^2}\; 
\biggl(\frac{x}{z} \biggr)
\biggl[
4\biggl(1-\frac{x}{\zeta}\biggr) + \frac{\zeta-2x}{z}
\biggr]\;G_{\zeta}(z,\mu)
\biggr\}
\eea


\bea
\label{eq:a6}
\mu\frac{\partial}{\partial \mu}G_{\zeta}(x,\mu)
&=&
\frac{\alpha_S(\mu)}{\pi} C_F\;
\biggl\{
\int\limits_0^x dz\; \biggl(\frac{\xprime}{\zprime} \biggr)
\biggl(1-\frac{x}{\zeta} \biggr)\; \Sigma_{\zeta}(z,\mu)
+
\int\limits_x^1 dz \biggl(2-\frac{x^2}{z \zeta} \biggr)\;
\Sigma_{\zeta}(z,\mu)
\biggr\}
\nonumber \\
\nonumber \\
&+&
\frac{\alpha_S(\mu)}{\pi} N_c\;
\biggl\{
\int\limits_0^x dz\;  \biggl(\frac{\xprime}{\zprime} \biggr)\;
\biggl[
\frac{2}{\zeta} \biggl(1-\frac{x}{\zeta} \biggr)
\biggl(1+2\frac{x}{\zeta}+\frac{x}{\zeta-z} \biggr)\; G_{\zeta}(z,\mu) 
\nonumber \\
\nonumber \\
& &\qquad\qquad\quad +\;
\frac{(\xprime/\zprime)\; G_{\zeta}(z,\mu)-G_{\zeta}(x,\mu)}{x-z}
\biggr]
\nonumber \\
\nonumber \\ 
&+&
\int\limits_x^1 dz\;  \biggl(\frac{x}{z} \biggr)
\biggl[
\frac{2x}{\zeta^2} 
\biggl(3-2\frac{x}{\zeta}+\frac{\zeta-x}{z} \biggr)\; G_{\zeta}(z,\mu) 
+
\frac{(x/z) G_{\zeta}(z,\mu)-G_{\zeta}(x,\mu)}{z-x}
\biggr]
\nonumber \\
\nonumber \\
& &\qquad\qquad\quad +\;
G_{\zeta}(x,\mu)\;
\biggl[ \frac{11-2/(3N_f)}{2N_c}+\ln\frac{x(1-x)}{\zeta}
\biggr]
\biggr\}~.
\eea
For $\zeta=1$ the above equations reduce to the ERBL evolution equations
for the distribution amplitudes. It is also instructive to check that
both set of equations,  
that is (\ref{eq:a3},\ref{eq:a4}) and (\ref{eq:a5},\ref{eq:a6}), 
lead to the same limiting set of equations
when $x \rightarrow \zeta$ from both sides.

The four equations (\ref{eq:a3}-\ref{eq:a6})
form a coupled set of equations which, in general, need to be solved
simultaneously. However for  $x>\zeta$ it is enough
to solve the first two  since the  right hand side of these equations
involves parton distribution for values $z>x$ (as is true for the
DGLAP equations in the limit $\zeta=0$).
This is not the case if $x<\zeta$. The  solution 
depends on values of parton distributions in the full interval $(0,1)$ and
we have to solve all four equations together.


\end{document}